\documentclass[twocolumn,showpacs,preprintnumbers,draftclsnofoot]{revtex4}
\usepackage{amsmath}
\usepackage{amssymb}
\usepackage{graphicx}
\usepackage{dcolumn}
\usepackage{bm}
\usepackage{amssymb}
\usepackage{graphicx}
\usepackage{dcolumn}
\usepackage{bm}
\begin{document}

\title{ Localized Refractive index sensing by integrated photonic crystal waveguide with edge-cavity }
\author{Ma Luo\footnote{Corresponding author:luoma@gpnu.edu.cn}, Kaichan Zhong, Jieli Luo }
\affiliation{School of Optoelectronic Engineering, Guangdong Polytechnic Normal University, Guangzhou 510665, China}

\begin{abstract}

We have theoretically proposed a highly compact refractive-index sensor consisted of edge-cavity and line-defect waveguide in two-dimensional photonic crystal. The sensing object is completely outside of the single enclosed surface of the sensor. The edge-cavity is designed by engineering the spatial distribution of the cutoff frequency of edge modes. The coupling between the edge-cavity and the waveguide is maximized by optimizing the radius of the rods between them, so that the transmittance spectrum through the waveguide has a sharp anti-peak. As the refractive index of the sensing object changes, the resonant wavelength of the edge-cavity is changed, which in turn changes the wavelength of the anti-peak. The sensitivity of the sensor is up to 40 nm/RIU, and the footprint of the sensor is only 40 $\mu m^{2}$. Because the transmittance spectrum is determined by the overlap between the sensing object and the highly localized resonant mode, the sensor can also perceive spatial distribution of refractive index in the sensing object.

\end{abstract}

\maketitle

\section{Introduction}

Refractive index sensing of medium implemented by optical measurement of transmittance and reflectance provides high speed detection of environmental changes \cite{Pevec18,Khaliji20}, such as temperature\cite{Qiu12,Kavungal18}, humidity\cite{XWang16} and chemical composition\cite{Weiss16,HLu18,YXu19}. The sensing mechanism is based on the measurement of the optical transmittance or reflectance spectrum, which is dependent on the refractive index of the sensing object. Various types of optical resonance have been proposed to enhance the sensitivity\cite{Tazawa07,DKCWu09,YliTYang13,JLiRYu16,YZhang18,BLiu18,BFWan21,Kostyukov21,Gryga21}, such as surface plasmon polaritons (SPPs) and localized surface plasmon resonances (LSPRs) \cite{Rivas04,Piliarik09,JZhangL12} in metallic nanostructure, bound states in the continuum (BICs) in dielectric grating, resonant nano-cavity in photonic crystal, and photonic crystal fiber.

For SPPs and LSPRs, the optical field is highly localized at the interface between dielectric and metal, so that the resonant frequency is highly sensitive to the refractive index of the dielectric background\cite{Henzie07,Chamanzar20,Joseph20,Becker10,WLiXJiang15,LShiJShang20}. Doped graphene can also host SPPs and LSPRs, so that nanostructure consisted of graphene can also enhance the sensitivity against the refractive index\cite{YXiao14,LSunYZhang19,SChen22}. Because the absorption losses of metal and graphene at visible and near-infrared wavelengths are large, the Q factor of the SPPs and LSPRs are usually low, which limit the figure of merit (FoM) of the sensor. As a result, sensing devices that is fully consisted of dielectric materials are more attractive.

BICs in dielectric grating theoretically have infinitely large Q factor\cite{CWHsu16,Koshelev19,Sadreev21,SJoseph21}. By tuning the parameter of the dielectric nanostructure, the BICs are transformed into quasi-BICs, which have ultra-large Q factor. Refractive index sensing devices based on quasi-BICs with large sensitivity as well as FoM have been proposed\cite{Srivastava19,Romano18,YWangMA18,Romano19,TCTanYK21,JWangJKuhne21,ZLiYXiang22,TSangSA21,Maksimov22,fengwu22}. Theoretical analysis of quasi-BICs are usually based on strictly periodic structures with infinite period, which is excited by incident plane wave. For realistic devices, hundreds of periods of the grating are required to avoid the finite size effect. Incident Gaussian beam with beam-width being a few times larger than the footprint of the device is required to simulate the plane wave incidence. As a result, the sensing device cannot be compactly integrated with photonic circuit. Integrated photonic has been proposed to be an upgraded version of integrated circuit \cite{Marpaung19,Shastri21}. Integration of multiple types of functional components in one integrated photonic circuit is essential for building applicable photonic product. Thus, designing refractive index sensors that can be compactly integrated with photonic circuit is necessary.

Refractive index sensors consisted of resonant cavity in photonic crystal have high sensitivity \cite{Hosseinzadeh21}, because the mode pattern of the resonant cavity completely overlaps with the sensing object (the background medium of the photonic crystal). Meanwhile, the measurement of the transmittance spectrum is through the input and output waveguides, so that the sensors can be compactly integrated with photonic circuit. However, the respondent speed of the sensors would be slow due to the following reason. The sensing object is the medium (gas or liquid) at the background of the photonic crystal. As a second medium (with different refractive index) appear outside of the sensor, the periodic rods of the photonic crystal could slow down the diffusion of the second medium into the background of the photonic crystal. The sensors cannot generate reliable prediction of the refractive index until the second medium become uniform background medium of the photonic crystal after long diffusion. Thus, the sensors cannot have real-time response in scene with flowing gas or liquid. Photonic crystal fiber sensors can also obtain high sensitivity \cite{ZheZhang19}, but the respondent speed of the sensor would also be slow due to the same reason. As a result, a refractive index sensor consisting of solid structure with the sensing object being outside of a smooth single enclosed surface is demanded.

In this paper, we propose a refractive index sensor, which has small footprint near to the surface of a photonic crystal with flat single enclosed surface, and measures transmittance spectrum of input-output waveguide as sensing signal. The device is based on the semi-infinite photonic crystal, which is consisted of two-dimensional square lattice of $Si$ rods in $SiO_{2}$ substrate. The structure of the device includes three parts, which are a line-defect waveguide inside the photonic crystal, an edge-cavity at the surface of the photonic crystal, and a coupling cavity between the waveguide and the edge-cavity. The coupling between the waveguide and the edge-cavity is optimized, so that the transmittance spectrum has a sharp anti-peak. The mode pattern of the edge-cavity partially overlaps with sensing object, such as gas or liquid, outside of the top surface of the semi-infinite photonic crystal, so that the resonant frequency of the edge-cavity is dependent on the refractive index of the sensing object. Because the anti-peak of the transmittance spectrum is nearly the same as the resonant frequency of the edge-cavity, the transmittance spectrum is sensitive to the refractive index of the sensing object. The mode pattern of the edge-cavity is highly localized, so that the sensor could identify sensing object with non-uniform refractive index, such as a droplet on the surface of the sensor.

This paper is organized as follows. In Sec. II, we describe the structure of the edge-cavity and optimized structure of the coupling cavity between the edge-cavity and the waveguide. In Sec. III, the sensing performance of the devices is illustrated, including the sensitivity and figure of merit for sensing homogeneous medium, and the sensing ability of droplet at the surface of the device. Finally, the conclusion is given in Sec. IV.

\section{Structure and Theoretical model}

The structure of the sensing device is based on a semi-infinite two-dimensional photonic crystal in lower half-space, which is consisted of square lattice of dielectric rods in dielectric substrate. The substrate is made of $SiO_{2}$ with refractive index being 1.46, and the dielectric rods are made of $Si$ with refractive index being 3.4. The period of the lattice is $L_{x}=L_{y}=500$ nm, and the radius of the dielectric rod is $R=0.225L_{x}$. The distance between the top surface of the semi-infinite photonic crystal and the center coordinate of the rods in the top row is equal to $L_{y}/2$. The sensing object, which could be gas or liquid, is the background medium above the top surface with refractive index being designated as $n_{s}$. Because both substrate and rods of the photonic crystal are made of solid with the top surface being the enclosed surface of the sensor, the sensing object cannot penetrate into the structure of the sensor. The sensing object is firstly assumed to be air with $n_{s}=1$.

In order to design refractive index sensor with import and export waveguide mode, the Maxwell equations is numerically simulated by applying the finite difference frequency domain (FDFD) method \cite{YenChungChiang09}. At the boundary of the computational domain, the perfectly match layer (PML) is applied to simulate the scattering of the electromagnetic wave. At the physical-domain boundary that cut through the import waveguide, the total field-scattering field boundary condition is applied to simulate the import of waveguide mode. The transmittance to a surface can be calculated by integrating the exiting component of the Poynting vector at the surface. For bulk photonic crystal with the same parameters, the band structure of the TM mode have a band gap with frequency range being $[0.2566,0.30921]\times2\pi c/L_{x}$. Thus, a point-defect or line-defect inside the photonic crystals could host micro-cavity modes or waveguide modes with resonant frequency within the gap. As a result, the numerical calculation is limited to TM mode.

\subsection{Edge-cavity}

\begin{figure}[tbp]
\scalebox{0.63}{\includegraphics{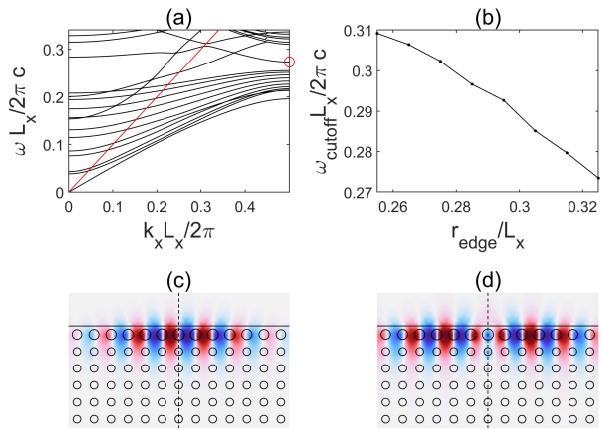}}
\caption{ (a) Band structure of semi-infinite photonic crystal with the radius of the dielectric rod at the edge being enlarged to be $0.325L_{x}$. The red line represents the light cone, and the red circle dot marks the cutoff frequency of the surface states. (b) Cutoff frequency of the surface states versus the radius of the dielectric rod at the edge of the semi-infinite photonic crystal. Structure of the edge-cavity at the surface of the semi-infinite photonic crystal are plotted as the solid lines in (c) and (d). In (c) and (d), the electric field patterns, $Re[E_{z}]$, of the two resonant modes with resonant frequency being $0.2794\times2\pi c/L_{x}$ and $0.2897\times2\pi c/L_{x}$ are plotted, respectively. }
\label{figure_cavity}
\end{figure}

Edge cavity can be engineered by modifying the radius of the dielectric rods at the surface of the semi-infinite photonic crystal. If the radius of all dielectric rods at the surface are enlarged to be $0.325L_{x}$, a band of surface modes appear within the band gap of the photonic crystal, and under the light cone \cite{LingLu12,Robertson93,RDMeade91}, as shown in Fig. \ref{figure_cavity}(a). At $k_{x}L_{x}/2\pi=0.5$, the frequency of the band of the surface mode reaches minimum, which is the cutoff frequency. The cutoff frequency is dependent on the radius of the dielectric rods at the surface, as shown in Fig. \ref{figure_cavity}(b). If the cutoff frequency at the left and right part of the surface is larger than that at the middle part of the surface, the surface mode could be trapped in the middle part of the surface. The mechanism of the surface mode trapping is similar to that of the $Fabry-P\acute{e}rot$ cavity. At the middle part (left and right parts) of the surface, the optical field can (cannot) travel along the surface, because the cutoff frequency is lower (higher) than the frequency of the optical field. As the surface traveling wave in the middle part of the surface reaches the left or right part of the surface, the wave is reflected back to the middle part of the surface. For certain frequency, the surface traveling mode, which is reflected back and forth between the left and right parts of the surface, satisfies the $Fabry-P\acute{e}rot$ resonant condition. Thus, a localized resonant edge-cavity mode appears. Depending on the spatial distribution of the cutoff frequency, more than one localized resonant edge-cavity modes could satisfy the $Fabry-P\acute{e}rot$ resonant condition.

In our design, the radius of the dielectric rod at the surface is dependent on the horizontal coordinate $x$ of each dielectric rod as $R_{surface}(x)=R(1+\delta e^{-(x-x_{c})^{2}/d_{c}^{2}})$, with $x_{c}$ being the coordinate of the center of the edge-cavity. The structure of the edge-cavity has mirror symmetry with the symmetric plane being marked by the vertical dashed line in Fig. \ref{figure_cavity}(c,d). Assuming $\delta=0.1$ and $d_{c}=8L_{x}$, two edge-cavity modes with resonant frequencies being $0.2794\times2\pi c/L_{x}$ and $0.2897\times2\pi c/L_{x}$ are found, whose mode pattern are odd and even symmetric about the symmetric plane, as shown in Fig. \ref{figure_cavity}(c) and (d), respectively. The Q factor of the two resonant modes are $2.9\times10^{7}$ and $3.6\times10^{5}$, respectively. The energy loss of the edge-cavity modes is due to scattering of the surface traveling wave to the plane wave above the light cone. The scattering mainly occurs at the interface between the middle and the left(right) parts of the surface. The first edge-cavity mode is more localized to the center of the cavity than the second resonant mode. Thus, the first edge-cavity mode is less overlapped with the interface than the second resonant mode, so that the Q factor of the first resonant mode is larger than that of the second resonant mode. For cavity with larger size (larger value of $d_{c}$), the edge-cavity modes become less overlapped with the interface, so that the Q fact can be enlarged. The evanescent wave of the mode pattern penetrate into the sensing object. Thus, as the refractive index of the sensing object changes, the resonant frequency of the edge-cavity modes change. By choosing different parameter $\delta$ and $d_{c}$, or constructing different form of the function $R_{surface}(x)$, varying types of edge cavities can be designed. More generally, edge-cavity can be engineered in photonic crystal with varying type of lattice structure, such as triangular lattice, as long as a large part of the band gap in the semi-infinite lattice being below the light cone.

\subsection{Coupling between edge-cavity and line-defect waveguide }

\begin{figure}[tbp]
\scalebox{0.63}{\includegraphics{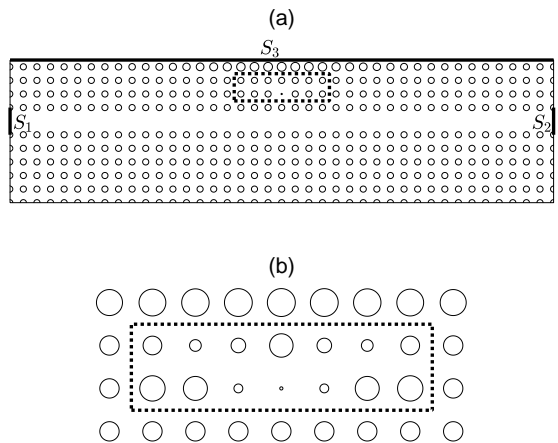}}
\caption{ (a) Structure of the integrated photonic crystal waveguide and edge-cavity, which are coupled by a micro-cavity within the dashed-rectangular between them. (b) The optimized structure of the coupling micro-cavity. From left to right, the radius of the rods of the top row are scaled by the factor of 0.9660, 0.6049, 0.7535,   1.2120, 0.7535, 0.6050, 0.9661, and those of the bottom row are scaled by the factor of 1.3106, 1.2396, 0.4577, 0.1633, 0.4577, 1.2397, 1.3106, relative to $0.225L_{x}$.  }
\label{figure_structure}
\end{figure}

The basic structure of the sensing device is given in Fig. \ref{figure_structure}(a). A line-defect waveguide is constructed by removing one row of dielectric row inside the photonic crystal. The waveguide and the edge-cavity are separated by three rows of dielectric rods. In order to increase the coupling strength between the waveguide and the edge-cavity, a coupling cavity is constructed by shrinking the radius of one dielectric rod between the waveguide and the edge-cavity. The radius of the dielectric rods at the center of the coupling cavity is optimized, so that the resonant frequency of the coupling cavity equates to that of the edge-cavity. The imported and exported optical wave through the waveguide mode occur at the boundaries marked as $S_{1}$ and $S_{2}$ in Fig. \ref{figure_structure}(a), respectively, which can be connected to photonic circuit. Part of the imported optical wave is scattered to exit the top surface, which is marked as $S_{3}$ in Fig. \ref{figure_structure}(a). Total field-scattering field boundary condition is applied at $S_{1}$ to simulate the import of optical wave to the waveguide mode. By calculating the total energy flux through the three surfaces, designated as $P_{1,2,3}$ through $S_{1,2,3}$, the transmittance spectrum from the import to the export and to the top surface are given as $S_{12}=P_{2}/P_{1}$ and $S_{13}=P_{3}/P_{1}$, respectively.

\begin{figure}[tbp]
\scalebox{0.39}{\includegraphics{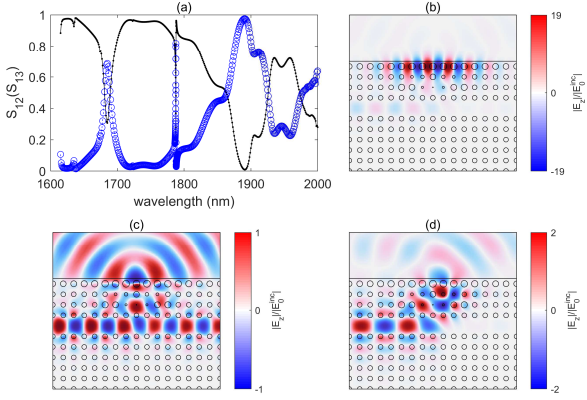}}
\caption{ (a) Transmittance to the $S_{2}$ and $S_{3}$ ports versus incident wavelength of the waveguide mode imported from $S_{1}$ are plotted as black (solid) and blue (empty) dots, respectively. (b-d) Distribution of electric field with incident wavelength being 1787.7 nm, 1685 nm, and 1891 nm, which corresponding to incident frequencies being $0.2797\times2\pi c/L_{x}$, $0.2967\times2\pi c/L_{x}$, and $0.2644\times2\pi c/L_{x}$, respectively. The field pattern is normalized by the amplitude of the incident field $E_{0}$.  }
\label{figure_pattern}
\end{figure}

Near to the resonant frequency $0.2792\times2\pi c/L_{x}$ of the first edge-cavity model, an anti-peak of $S_{12}$ at frequency $0.2797\times2\pi c/L_{x}$ are found. The width of the anti-peak is small, but the minimum transmittance at the anti-peak is around 0.7, which imply that the coupling between the waveguide and the edge-cavity is weak. The weakness of the coupling is due to the mismatching between the mode pattern of the edge-cavity and the coupling cavity. By further optimizing the radius of the 14 dielectric rods near to the coupling cavity [the rods surrounded by the dashed rectangular in Fig. \ref{figure_structure}(a)], the coupling strength is maximized. The optimization is implemented by using numerical Newton-Raphson method with the radius of the 14 dielectric rods being arguments, and the transmittance at the anti-peak being objected function. The result of the optimization is plotted in Fig. \ref{figure_structure}(b), along with the numerical result of the scaling factor of the radius of the 14 rods in the caption. For this structure, the transmittance spectrum of $S_{12}$ and $S_{13}$ are plotted in Fig. \ref{figure_pattern}(a). The minimum transmittance at the anti-peak of $S_{12}$ at frequency $0.2797\times2\pi c/L_{x}$ is as small as 0.18. The optimized structure mixed the edge-cavity mode with the coupling cavity mode, so that the Q factor of the edge-cavity is decreased. As a result, the width of the anti-peak is 0.4 nm, which is much larger than the line-width corresponding to the Q factor of the resonant edge-cavity mode in the absence of the coupling cavity. Because the resonant mode of the edge-cavity is highly localized, the footprint of the sensing device is limited to the area in Fig. \ref{figure_pattern}(b), which is 40 $\mu m^{2}$. In addition, two more anti-peaks appear at frequencies $0.2967\times2\pi c/L_{x}$ and $0.2644\times2\pi c/L_{x}$.

The field pattern of electric field at the three anti-peak of $S_{12}$ are plotted in Fig. \ref{figure_pattern}(b-d). The mechanism of the two cases in Fig. \ref{figure_pattern}(b) and (c) are the same. The incident waveguide field excite one cavity mode, which induces radiation into the free space. Thus, the line-shape of the anti-peak are both Gaussian. For the case in Fig. \ref{figure_pattern}(b), the resonant mode of the edge-cavity is strongly excited, so that the amplitude of the electric field at the edge-cavity is 19 times larger than that of the incident field from the import at $S_{1}$. The field pattern at the coupling cavity is odd about the symmetric plane of the structure, whose symmetry matched with that of the edge-cavity mode. The oscillating period of the electric field along the axis of the waveguide is approximately two times larger than that along the horizontal axis of the edge-cavity. The optimized structure of the coupling cavity effectively transform the oscillating period of the electric field, so that the incident field from the waveguide can be strongly coupled to the edge-cavity. For the case in Fig. \ref{figure_pattern}(c), the incident field excited the resonant mode of the coupling waveguide, whose field pattern has even symmetry about the symmetric plane. The quality factor of the coupling cavity is low, so that the amplitude of the electric field is not highly enhanced, and the width of the corresponding anti-peak of $S_{12}$ is large. Only 70 percents of the incident power is radiated through $S_{3}$, and 30 percents of the incident power is transmitted to the export at $S_{2}$. For the case in Fig. \ref{figure_pattern}(d), the combination of the coupling cavity and the edge-cavity functions as optical wave redirecting device, so that the incident power is completely radiated through the surface. Because no resonant mode is excited, the line-shape of the anti-peak is not Gaussian.

\section{Sensing performance}

\begin{figure}[tbp]
\scalebox{0.38}{\includegraphics{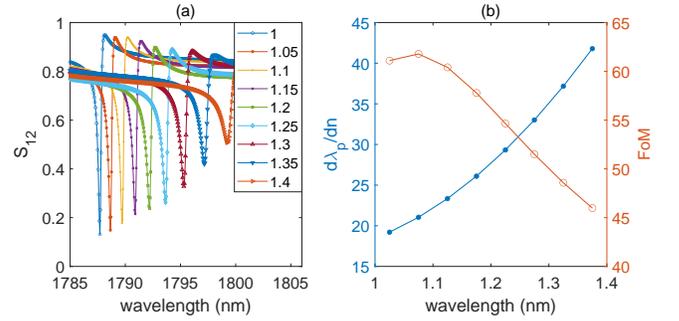}}
\caption{ (a) Transmittance to the $S_{2}$ port versus wavelength around the resonant peak at 1787.7 nm for varying value of air refractive index $n_{s}$, whose value is given in the legend. (b) The sensitivity of the peak versus $n_{s}$ and the FoM, which are extracted from (a), are plotted as solid and empty dotted lines, respectively.   }
\label{figure_sensitivity}
\end{figure}

We firstly consider the case with $n_{s}$ being uniform. As $n_{s}$ changes, the resonant frequency of the edge-cavity changes, which in turn changes the transmittance spectrum of the anti-peak corresponding to the resonant mode in Fig. \ref{figure_pattern}(b). As a result, the anti-peak of the transmittance spectrum shifts to larger wavelength, as shown in Fig. \ref{figure_sensitivity}(a). Meanwhile, the minimum transmittance and the line-width of the anti-peak increases, because the coupling strength between the edge-cavity and the coupling cavity become weaker. However, the correlation between the transmittance spectrum and the refractive index of the sensing object is still large. As $n_{s}$ change from 1 to 1.05, the frequency of the anti-peak shifts from 1787.7 nm to 1788.7 nm. As a result, the sensitivity is 20 nm/RIU. As $n_{s}$ increases to 1.4, the sensitivity reaches up to 40 nm/RIU, as shown in Fig. \ref{figure_sensitivity}(b). In addition to sensitivity, the FoM is calculated as $F=\frac{S}{\Delta\lambda}$, where $S$ is the sensitivity and $\Delta\lambda$ is the line width of the anti-peak. As $n_{s}$ increases from 1 to 1.4, the FoM decreases from 60 to 45. Although the sensitivity of the proposed sensor is smaller than that of the other type of sensor, which is usually up to 400 nm/RIU \cite{Hosseinzadeh21}, the structure of the proposed sensor has single enclosed surface. Thus, the sensor does not mechanically interfere the flowing motion of the sensing object.

\begin{figure}[tbp]
\scalebox{0.58}{\includegraphics{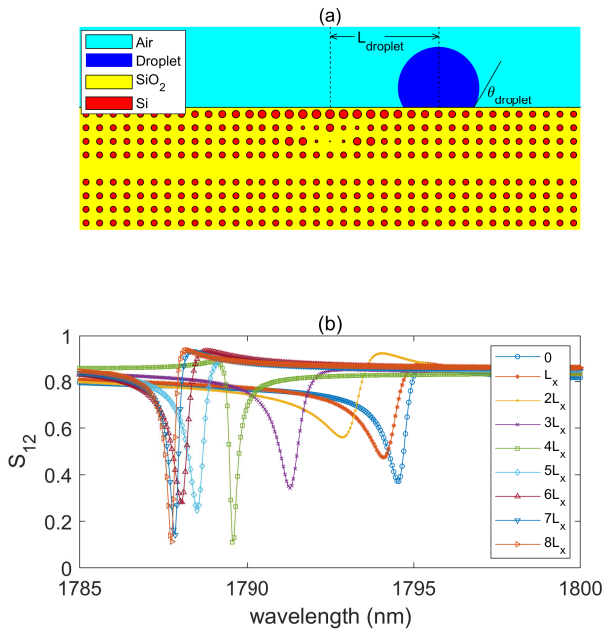}}
\caption{ (a) Scheme of sensing horizontal location, $L_{droplet}$, of the liquid droplet on the surface of the photonic crystal. (b) Transmittance versus incident wavelength for varying $L_{droplet}$, whose value is given in the legend.   }
\label{figure_droplet}
\end{figure}

Secondly, we consider the case with $n_{s}$ being nonuniform. Because the resonant mode pattern of the edge-cavity is highly localized within a small region, the transmittance is mainly sensitive to the refractive index of the sensing object near to the edge-cavity. As a result, the device can be applied to sense local change of the refractive index. Specifically, the localized region of the resonance mode pattern of the edge-cavity in Fig. \ref{figure_pattern}(b) is limited within the horizontal coordinate as $x\in[x_{c}-3L_{x},x_{c}+3L_{x}]$. Thus, the spatial resolution of the sensing device could be around $L_{x}$. One example of the application is illustrated in Fig. \ref{figure_droplet}(a), where a droplet of liquid is attached on the flat surface of the $SiO_{2}$ substrate. For the coordinate outside and inside of the droplet, $n_{s}(x,y)$ is equal to one and $n_{droplet}$, respectively, with $n_{droplet}$ being the refractive index of the liquid. The structure of the droplet is characterized by radius and infiltration angle, which are designated as $R_{droplet}$ and $\theta_{droplet}$, respectively. The horizontal distance between the center of the droplet and $x_{c}$ is designated as $L_{droplet}$. Assuming $n_{droplet}=1.3$, $R_{droplet}=3L_{x}$, and $\theta_{droplet}=30^{o}$, the transmittance spectrum of varying $L_{droplet}$ are plotted in Fig. \ref{figure_droplet}(b). As $L_{droplet}=0$, the droplet completely covers the localized region of the resonance mode pattern of the edge-cavity, so that the transmittance spectrum is mainly determined by the refractive index of the droplet. Thus, the transmittance spectrum is similar to the case in Fig. \ref{figure_sensitivity}(a) with $n_{s}=1.3$. As the droplet moves away from the edge-cavity, the boundary of the droplet cut across the localized region of the resonant mode pattern of the edge-cavity. As $|L_{droplet}|$ increases, the average refractive index within the localized region of the resonant mode pattern of the edge-cavity decreases, so that the wavelength of the anti-peak of the transmittance spectrum monotonously decreases, as shown in Fig. \ref{figure_droplet}(b). Because of the scattering of the evanescent wave at the boundary of the droplet, the line width and wavelength of the anti-peak of the transmittance spectrum are not linearly dependent on $L_{droplet}$. For droplet with varying parameters(including $n_{droplet}$, $R_{droplet}$, $\theta_{droplet}$, and $L_{droplet}$), a more comprehensive inverse problem solver, such as deep learning\cite{LFengValenza22,Orazbayev20}, could be applied to speculate the parameters according to the measurement of the transmittance spectrum.

\section{Conclusion}

In conclusion, we theoretically propose a refractive index sensor based on edge-cavity and line-defect waveguide in two-dimensional dielectric photonic crystal. The structure of the sensor has flat enclosed surface, which does not interfere with the flowing of the sensing object outside of the surface. The coupling between the line-defect waveguide and the edge-cavity is optimized by designing the structure of the coupling cavity between them. Utilizing the high Q-factor and localization of the resonant mode of the edge-cavity, the photonic structure with small footprint achieves high-performance refractive-index sensing. The sensitivity is as large as 40 nm/RIU, and the FoM reaches 60. The sensing device could perceive the change of spatial distribution of refractive index within a small region near to the edge-cavity, such as sensing the location of a droplet on the surface of the substrate. This type of sensing device with small footprint could have potential application in sensing bio-solution with vesicles or cells.

\begin{acknowledgments}
This project is supported by the Natural Science Foundation of Guangdong Province of China (Grant No.
2022A1515011578), the Project of Educational Commission of Guangdong Province of China (Grant No. 2021KTSCX064), and the Startup Grant at Guangdong Polytechnic Normal University (Grant No. 2021SDKYA117).
\end{acknowledgments}

\clearpage

\end{document}